\documentclass[12pt]{article}
\usepackage{epsfig}

\def\beq{\begin{equation}}
\def\eeq#1{\label{#1}\end{equation}}
\def\eeqn{\end{equation}}

\def\beqa{\begin{eqnarray}}
\def\eeqa#1{\label{#1}\end{eqnarray}}
\def\eeqan{\end{eqnarray}}

\let\bar=\overbar

\def\Dslash{\not{\hbox{\kern-4pt $D$}}}
\def\dslash{\not{\hbox{\kern-2pt $\del$}}}

\def\msb{{\bar{\ssstyle M \kern -1pt S}}}

\usepackage{fancyhdr,graphicx}
\def\Title#1{\begin{center} {\Large {\bf #1} } \end{center}}

\begin{document}

\Title{Induced Magnetism in Color-Superconducting Media}

\bigskip\bigskip


\begin{raggedright}

{\it Efrain J. Ferrer\index{Med, I.}\\
Department of Physics\\
The University of Texas at El Paso\\
El Paso\\
TX 79968\\
USA\\
{\tt Email: ejferrer@utep.edu}}
\bigskip\bigskip
\end{raggedright}

\begin{abstract}
The dense core of compact stars is the natural medium for the realization of color superconductivity. A common characteristic of such astrophysical objects is their strong magnetic fields, especially those of the so called magnetars. In this talk, I discuss how a color superconducting core can generate or/and enhance the stellar magnetic field independently of a magnetohydrodynamic dynamo mechanism. The magnetic field generator is in this case a gluonic current which circulates to stabilize the color superconductor in the presence of a strong magnetic field or under the pairing stress produced in the medium by the neutrality and $\beta$-equilibrium constraints.
\end{abstract}

\section{Introduction}

The extremely high densities reached in the cores of compact stars make those astrophysical objects the natural place for the realization of color superconductivity.
If the core matter density is high enough to guarantee the quark deconfinement, but
sufficiently low to consider the decoupling of the strange quark
mass $M_s$, the quarks can condense to form a two-flavor color
superconductor (2SC) \cite{Bailin-Love}. At higher densities, for
which $M_s$ cannot be regarded as extremely large, the three-flavor
phases $gCFL$ \cite{Alford} and $CFL$ \cite{alf-raj-wil-99/537} will
appear respectively with increasing densities. Matter inside compact
stars should be neutral and remain in $\beta$-equilibrium. When
these conditions along with $M_s$ are taken into account, a pairing mismatch takes place which is reflected in the
dynamics of the gluons in the $gCFL$ phase (or for that matter in
the 2SC phase if the s-quark is decoupled). As a consequence, some gluon modes become
tachyonic \cite{Igor,Fukushima}, indicating that the system ground
state should be restructured.

The chromomagnetic instabilities created by pairing mismatch has led to several interesting proposals.
Some of the most promising possibilities are a modified $CFL$-phase
with a condensate of kaons \cite{Schafer} that requires a
space-dependent condensate to remove the instability; a LOFF phase
on which the quarks pair with nonzero total momentum \cite{LOFF};
and an inhomogenous gluon condensate together with the spontaneous induction of an inmedium magnetic field \cite{Ind-Vortex}. At present
it is not clear if any of these proposals, or a combination of them, is the final solution to
the instability problem.

The phase with a ground state formed by
an inhomogeneous condensate of charged gluons and induced rotated
magnetic field was found in Ref. \cite{Ind-Vortex} in the Meissner unstable
region of the so-called gapped $2SC$ phase \cite{Huang}, but it is
expected to be also realized in the three-flavor theory. It
has the peculiarity of preserving the electromagnetic gauge
invariance $\widetilde{U}_{em}(1)$ of the color superconductor.

On the other hand, a common characteristic of dense astrophysical objects is their strong magnetic fields, especially those found for the so called magnetars which can range between $10^{14} - 10^{16}$ G \cite{magnetars} at their surface.
The stars' interior having a larger density can reach even higher values thanks to the magnetic flux conservation in the stellar medium. Maximum strengths of $10^{18}-10^{19}$ G are allowed
by a simple application of the virial theorem \cite{Lai}.

As found in Ref. \cite{Vortex}, the magnetic field can also
influence the gluon dynamics as we will discuss below. At field strengths
comparable to the charged gluon Meissner mass a chromomagnetic instability is created which is removed by the realization of a new phase which is formed
by the generation of an inhomogeneous condensate of
$\widetilde{Q}$-charged gluons. The generated gluon condensate
anti-screens the magnetic field due to the anomalous magnetic
moment of these spin-1 particles. Because of the anti-screening,
this condensate does not give a mass to the $\widetilde{Q}$
photon, but instead amplifies the applied rotated magnetic field.
This means that at such applied fields the color superconductor behaves as a
paramagnet.

The capability exhibited by color superconductors to generate or enhance a magnetic field due to the gluon-vortex antiscreening
mechanism can be of interest for
astrophysics, since compact stars with color superconducting cores could
have larger magnetic fields than neutron stars made up entirely of
nuclear matter.

As follows, I will discuss the role of gluons in generating and/or enhancing magnetic fields in color superconductors.
\section{Charged Gluons in Color Superconductivity}
In spin-zero color superconductivity, the color condensates in the CFL, as well as in the 2SC phases,
have non-zero electric charge. Then, we could expect a magnetic response of the color superconducting medium similar to that of the conventional superconductor, where the Cooper pairs are
electrically charged and consequently the electromagnetic gauge invariance is
spontaneously broken. In that case, the photon acquires a Meissner mass thus having the possibility
to screen a weak magnetic field (the well known phenomenon of Meissner effect). Nevertheless, in the spin-zero color superconductor the conventional electromagnetic field $A_\mu$ is not an eigenfield, but it is mixed with the $8^{th}$-gluon $G^8_\mu$. Hence, as in the electroweak model after the symmetry breaking produced by the condensation on the Higgs field, where the $SU(2)$ ($W_\mu^3$) and $U(1)$ ($B_\mu$) bosons combine to give rise to the real eigenfields, i.e. the $Z_\mu$ boson and $A_\mu$ photon, in the color superconductor it is the linear combinations of $A_\mu$ and  $G^8_\mu$ what becomes the in-medium physical modes \cite{alf-raj-wil}. In this case, one of the two linear combinations between $A_\mu$ and  $G^8_\mu$
\begin{equation}
\widetilde{A}_{\mu}=\cos{\theta}\,A_{\mu}-\sin{\theta}\,G^{8}_{\mu}
\ , \label{1}
\end{equation}
becomes massless and plays the role of the electromagnetic field inside the color superconducting medium (it is called the rotated electromagnetic field), while the orthogonal combination
\begin{equation}
\widetilde{G}_{\mu}^8=\sin{\theta}A_{\mu}+\cos{\theta}\,G^{8}_{\mu}
\ , \label{2}
\end{equation}
is massive. The mixing angle $\theta$ is a function of the strong $g$ and electromagnetic $e$ coupling constants, and depends on the nature of the color superconducting phase. In particular, for the CFL phase
\begin{equation}
\cos{\theta_{CFL}}=\frac{1}{\sqrt{1+\frac{4}{3}(\frac{e}{g})^2}},\,\quad \sin{\theta_{CFL}}=\frac{1}{\sqrt{1+\frac{3}{4}(\frac{g}{e})^2}}
\ , \label{angle-CFL}
\end{equation}
and for the 2SC phase it is given by
\begin{equation}
\cos{\theta_{2SC}}=\frac{1}{\sqrt{1+\frac{1}{3}(\frac{e}{g})^2}},\,\quad \sin{\theta_{2SC}}=\frac{1}{\sqrt{1+3(\frac{g}{e})^2}}
\ , \label{angle-2SC}
\end{equation}
Because of the hierarchy between the two coupling constants, in both phases the mixing angle $\theta$ is
sufficiently small ($\sin{\theta}\sim e/g\sim1/40$). Thus, the
penetrating field in the color superconductor (i.e. the rotated photon) is mostly formed by
the photon with only a small gluon admixture. Because the new massless field plays the role of the "in-medium" photon in the color
superconductor, the propagation of light in the color
superconductor is different from that in an electric superconductor.

Although in QCD gluons are electrically neutral, on the color superconducting
background they can interact with the rotated electromagnetism acquiring
a $\widetilde{Q}$ charge. In units of $\widetilde{e} =
e \cos{\theta_{CFL}}$ the gluons charges in the CFL phase are
\begin{equation} \label{table-2SC}
\begin{tabular}{|c|c|c|c|c|c|c|c|c|}
  \hline
  $G_{\mu}^{1}$ & $G_{\mu}^{2}$ & $G_{\mu}^{3}$ & $G_{\mu}^{+}$ & $G_{\mu}^{-}$ & $I_{\mu}^{+}$ & $I_{\mu}^{-}$ & $\widetilde{G}_{\mu}^{8}$ \\
  \hline
  0 & 0 & 0 & 1 & -1 & 1 & -1 & 0 \\
  \hline
\end{tabular} \ .
\end{equation}
where it was introduced the notation for the charged fields
$G_{\mu}^{\pm}=\frac{1}{\sqrt{2}}[G_{\mu}^{4}\pm iG_{\mu}^{5}]$
and $I_{\mu}^{\pm}=\frac{1}{\sqrt{2}}[G_{\mu}^{6}\pm
iG_{\mu}^{7}]$.

In the 2SC phase the charges of the gluons in units of $\widetilde{e} =
e \cos{\theta_{2SC}}$ are
\begin{equation} \label{table-CFL}
\begin{tabular}{|c|c|c|c|c|c|}
  \hline
  $G_{\mu}^{1}$ & $G_{\mu}^{2}$ & $G_{\mu}^{3}$ & $K_{\mu}$ & $K_{\mu}^{\dag}$ &  $\widetilde{G}_{\mu}^{8}$ \\
  \hline
  0 & 0 & 0 & 1/2 & -1/2 &  0 \\
  \hline
\end{tabular} \ .
\end{equation}
where we used for the charged fields the doublet representation
\begin{eqnarray}  \label{charged-fields-2SC}
K_{\mu} \equiv \frac{1}{\sqrt{2}}\left(
\begin{array}{cc}
G_{\mu}^{(4)}-iG_{\mu}^(5)\\
G_{\mu}^{(6)}-iG_{\mu}^(7)
\end{array}
\right) \
\end{eqnarray}
The charged gluon fields
$G_{\mu}^{\pm}$, $I_{\mu}^{\pm}$ and $K_{\mu}^{\pm}$ can interact, through the
long-range field $\widetilde{A}_{\mu}$, with an applied external
magnetic field.
\section{Enhanced Magnetic Fields by Gluon Vortices}
An applied magnetic field can modify the color superconducting pair-condensate structure. As shown in Ref. \cite{MCFL}, the
color-superconducting properties of the three-flavor system are substantially
affected by the penetrating $\widetilde{B}$ field and as a
consequence, a new phase, called Magnetic Color Flavor Locked (MCFL)
phase \cite{MCFL}-\cite{Phases}, takes place. Now, the magnetic field effect in color superconductivity is not only restricted to the ground-state pairing, but it can also impact the gluon dynamics. As follow, I will discuss how a sufficiently strong magnetic field can produce in the CFL phase a new background formed by vortices of rotated charged gluons, which as a back reaction boost the applied magnetic field \cite{Vortex}.

To investigate the effect of the applied rotated magnetic field
$\widetilde{H}$ on the charged gluons, we should start from the effective action of the charged fields $G_{\mu}^{\pm}$ (the
contribution of the other charges gluons $I_{\mu}^{\pm}$ is similar)
\begin{eqnarray}
\label{Eff-Act-2} &\Gamma_{eff}^{c}=  \int d^{4}x
\{-\frac{1}{4}(\widetilde{f}_{\mu
\nu})^{2}-\frac{1}{2}|\widetilde{\Pi}_{\mu}G_{\nu}^{-}-\widetilde{\Pi}_{\nu}G_{\mu}^{-}|^{2}&
\nonumber
 \\
& -  [(m_{D}^{2} \delta_{\mu 0} \delta_{\nu 0}+ m_{M}^{2}
\delta_{\mu i} \delta_{\nu i})+ i\widetilde{e}\widetilde{f}_{\mu
\nu}] G_{\mu}^{+}G_{\nu}^{-}& \nonumber
 \\
 & +
\frac{g^2}{2}[(G^{+}_{\mu})^{2}(G^{-}_{\nu})^{2}-(G^{+}_{\mu}G^{-}_{\mu})^{2}]+\frac{1}{\lambda}G_{\mu}^{+}\widetilde{\Pi}_{\mu}\widetilde{\Pi}_{\nu}G_{\nu}^{-}
\},&
\end{eqnarray}
where $\lambda$ is the gauge fixing parameter, $\widetilde{\Pi}_{\mu}=\partial_{\mu}
-i\widetilde{e}\widetilde{A}_{\mu}$ is the covariant derivative in the presence of the external rotated field, $m_{D}$ and $m_{M}$ are the $G_{\mu}^{\pm}$-field Debye and Meissner masses respectively, and the field strength tensor for the rotated electromagnetic field if denoted by $\widetilde{f}_{\mu
\nu}=\partial_{\mu}\widetilde{A}_{\nu}-\partial_{\nu}\widetilde{A}_{\mu}$.
 The corresponding Debye and Meissner masses in (\ref{Eff-Act-2})
are given by \cite{Wang}
\begin{equation}
m_{D}^{2} = m_{g}^{2} \frac{21-8 \ln 2}{18},\qquad m_{M}^{2} =
m_{g}^{2} \frac{21-8 \ln 2}{54},
\end{equation}
with $m_{g}^{2}=g^2(\mu^{2}/2\pi^{2})$. We
are neglecting the correction produced by the applied field to the
gluon Meissner masses since it will be a second order effect. The effective action (\ref{Eff-Act-2}) is characteristic of a
spin-1 charged field in a magnetic field (for details see for
instance \cite{emilio}).

Assuming that the penetrating magnetic field
points along the third spatial direction
($\widetilde{f}^{ext}_{12}=\widetilde{H}$), we find after diagonalizing the
mass matrix of the field components ($G^{+}_{1}, G^{+}_{2}$) in (\ref{Eff-Act-2})
\begin{equation}
\left(
\begin{array}{cc}
m_{M}^{2}& i\widetilde{e}\widetilde{H} \\
- i\widetilde{e}\widetilde{H}& m_{M}^{2}
 \label{mass-matrx}
\end{array} \right) \rightarrow
\left(
\begin{array}{cc}
m_{M}^{2}+\widetilde{e}\widetilde{H}& 0 \\
0& m_{M}^{2}-\widetilde{e}\widetilde{H}
 \label{mass-matrx}
\end{array} \right),
\end{equation}
with corresponding eigenvectors ($G^{+}_{1}, G^{+}_{2}$)
$\rightarrow$ ($G,iG$). We see that the gluon anomalous magnetic moment term $i\widetilde{e}\widetilde{f}_{\mu
\nu}G_{\mu}^{-}G_{\nu}^{+} $
produces for the lowest mass mode in (\ref{mass-matrx}) a sort of
"Higgs mass" above the critical field
$\widetilde{e}\widetilde{H}_{C}= m_{M}^2$, indicating that the
$G$-field grows exponentially with time (this is the well known "zero-mode
problem" found in the presence of a magnetic field for Yang-Mills
fields \cite{zero-mode}, for the $W^{\pm}_{\mu}$ bosons in the
electroweak theory \cite{Skalozub, Olesen}, and even for
higher-spin fields in the context of string theories
\cite{porrati}). Thus, it should be expected that the solution of
the instability is reached through the restructuring of the ground
state through the condensate of the field bearing the tachyonic mode (i.e. the $G$-field).

To find the $G$-field condensate and the induced magnetic field
$\widetilde{\textbf{B}}=\nabla\times\widetilde{\textbf{A}}$, with
$\widetilde{\textbf{A}}$ being the total rotated electromagnetic
potential in the condensed phase in the presence of the external
field $\widetilde{H}$, we should start from the Gibbs free energy
density $\mathcal{G}=\mathcal{F}-\widetilde{H}\widetilde{B}$, since
it depends on both $\widetilde{B}$ and $\widetilde{H}$
($\mathcal{F}$ is the system free energy density). Since
specializing $\widetilde{H}$ in the third direction the instability
develops in the $(x,y)$-plane, we make the ansatz $G=G(x,y)$.
Starting from (\ref{Eff-Act-2}) in the Feynman gauge $\lambda=1$,
which in terms of the condensed field $G$ implies
$(\widetilde{\Pi}_{1}+i\widetilde{\Pi}_{2})G=0$, we have that the
Gibbs free energy in the condensed phase is
\begin{eqnarray}
\label{Gibbs} \mathcal{G}_{c} =\mathcal{F}_{n0}
-2G^{\dag}\widetilde{\Pi}^{2}
G-2(2\widetilde{e}\widetilde{B}-m_{M}^{2})|G|^{2}+2g^{2}|G|^{4}\nonumber
 \\
+ \frac{1}{2}\widetilde{B}^{2}-\widetilde{H}\widetilde{B}\qquad
\qquad \qquad \qquad \qquad \qquad \qquad
\end{eqnarray}
where $\mathcal{F}_{n0}$ is the system free energy in the normal
phase ($G=0$) at zero magnetic field.

The equations of minimum for the fields $G$ and
$\widetilde{B}$ are respectively obtained from (\ref{Gibbs}) as
\begin{equation}
\label{G-Eq-1} -\widetilde{\Pi}^{2}
G+(m_{M}^{2}-2\widetilde{e}\widetilde{B})G+2g^{2}|G|^{2}G=0,
\end{equation}
and
\begin{equation}
\label{B-Eq-1} 2\widetilde{e} |G|^{2}-\widetilde{B}+\widetilde{H}=0
\end{equation}

Eqs. (\ref{G-Eq-1})-(\ref{B-Eq-1}) are analogous to the Ginsburg-Landau
equations for the conventional superconductor with $G$ playing the role of the complex order parameter. Nevertheless, there are two
distinctive factors that differentiate the Ginsburg-Landau equations of conventional superconductivity from (\ref{G-Eq-1})-(\ref{B-Eq-1}). They are given by the negative sign in front of the $\widetilde{B}$ field in Eq. (\ref{G-Eq-1}) and the positive sign in the first term of the LHS of Eq. (\ref{B-Eq-1}). The fact that here we get opposite signs to those appearing in conventional superconductivity is due to the different nature of the condensates in both cases. While in conventional superconductivity the Cooper pair is a spin-0 condensate, here we have a condensate formed by spin-1 charged particles which interact through their anomalous magnetic moment with the magnetic field (i.e. the term $i\widetilde{e}\widetilde{f}_{\mu
\nu}G_{\mu}^{-}G_{\nu}^{+} $ in (\ref{Eff-Act-2}) is responsible of that dissimilitude).

The positive sign in front of the first term in (\ref{B-Eq-1}) implies that the
condensation of the gluon field makes the magnetic field in the new
phase, $\widetilde{B}$, larger than the applied field,
$\widetilde{H}$. That is, the magnetic field is boosted to a higher
value which depends on the modulus of the $G$-condensate. Hence, the phase which is attained at $\widetilde{H}\geq \widetilde{H}_c$ is called paramagnetic CFL \cite{Vortex, Phases}. I want to point out that at the scale of baryon densities
typical of neutron-star cores ($\mu \simeq 400 MeV$, $g(\mu)\simeq
3$) the charged gluons magnetic mass in the CFL phase is $m_{M}^{2}
\simeq 16\times 10^{-3} GeV^{2}$. This implies a critical magnetic
field of order $\widetilde{H}_{c}\simeq 77\times 10^{16} G$. Although it is a significant high value, it is in the expected range
for the neutron star interiors. Let us stress that in our analysis
we considered asymptotic densities where quark masses can be
neglected (CFL phase).

To find the structure of the gluon condensate we should solve the non-linear differential equation (\ref{G-Eq-1}). However, to get an analytic solution we can consider the approximation where  $\widetilde{H}\approx \widetilde{H}_c=m_M^2$ and consequently $\mid G\mid \approx 0$. In this approximation, Eq. (\ref{G-Eq-1}) can be linearized as
\begin{equation}
\label{Vortex-Eq-1} [\partial_{j}^{2}-\frac{4\pi
i}{\widetilde{\Phi}_{0}}\widetilde{H}_{C}x\partial_{y}-4\pi^{2}\frac{\widetilde{H}_{C}^{2}}{\widetilde{\Phi}_{0}^{2}}x^{2}+\frac{1}{\xi^{2}}]G=0,
\quad j=x,y
\end{equation}
where we fixed the gauge condition
$\widetilde{A}_{2}=\widetilde{H}_{C}x_{1}$, and introduced the notations
$\widetilde{\Phi}_{0}\equiv2\pi/\widetilde{e}$, and
$\xi^{2}\equiv1/(2\widetilde{e}\widetilde{H}_{C}-m_{M}^{2})=1/m_{M}^{2}$.

Eq. (\ref{Vortex-Eq-1}) is formally similar to the Abrikosov's equation in type-II conventional superconductivity \cite{Abrikosov}, with $\xi$ playing the role of the coherence length and $\widetilde{\Phi}_{0}$ of the flux quantum per vortex cell. Then, following the Abrikosov's approach, a solution of Eq. (\ref{Vortex-Eq-1}) can be found as
\begin{equation}
\label{Vortex-solution} G(x,y)=\frac{1}{\sqrt{2}\widetilde{e}\xi}e^{-\frac{x^2}{e\xi^2}}\vartheta_3(u/\tau),
\end{equation}
with $\vartheta_3(u/\tau)$ being the elliptic theta function with arguments
\begin{equation}
\label{arguments} u=-i\pi b(\frac{x}{\xi^2}+\frac{y}{b^2}), \qquad \tau=-i\pi\frac{b^2}{\xi^2}
\end{equation}
In (\ref{arguments}) the parameter $b$ is the periodic length in the y-direction ($b=\Delta y$). The double periodicity of the elliptic theta function also implies that there is a periodicity length in the x-direction given by $\Delta x=\widetilde{\Phi}_{0}/b\widetilde{H}_{c}$. Therefore, the magnetic flux through each
periodicity cell ($\Delta x \Delta y$) in the vortex lattice is quantized $\label{Flux}
\widetilde{H}_c \Delta x \Delta y=\widetilde{\Phi}_{0}$, with
$\widetilde{\Phi}_{0}$ being the flux quantum per unit vortex cell.
In this semi-qualitative analysis we considered the Abrikosov's
ansatz of a rectangular lattice, but the lattice configuration should be carefully
determined from a minimal energy analysis. For the rectangular
lattice, we see that the area of the unit cell is $A=\Delta x \Delta
y=\widetilde{\Phi}_{0} /\widetilde{H}_c$, so decreasing with
$\widetilde{H}$. In summary, we have that to remove the instability
a magnetic field specialized along the $z$-direction turns
inhomogeneous in the $(x,y)$-plane since it depends on the
condensate $G$, which has a periodic structure on that plane, while it can be
homogeneous in the $z$-direction, therefore it forms a fluxoid along
the $z$-direction that creates a nontrivial topology on the
perpendicular plane. From (\ref{B-Eq-1}) we see that the magnetic
field can go from a minimum value $\widetilde{H}$ to a maximum at
the core of the fluxoid that depends on the amplitude of the gluon
condensate determined by the mismatch between the applied field and
the gluon Meissner mass.

Summarizing, at low $\widetilde{H}$ field, the CFL phase behaves as an
insulator, and the $\widetilde{H}$ field just penetrates through it.
At sufficiently high $\widetilde{H}$, the condensation of $G$
is triggered inducing the formation of a lattice of magnetic flux
tubes that breaks the translational and remaining rotational
symmetries. It should be noticed that contrary to the situation in conventional type-II
superconductivity, where the applied field only penetrates through the
flux tubes and with a smaller strength, the vortex state in the color superconductor has the peculiarity that outside the flux tube the
applied field $\widetilde{H}$ totally penetrates the sample, while
inside the tubes the magnetic field becomes larger than
$\widetilde{H}$.

\section{Induced Magnetic Field in Color Superconductivity with Chromomagnetic Instabilities}

As known, chromomagnetic instabilities can be present in color
superconductivity even in
the absence of an external magnetic field. As it was found in Ref. \cite{Igor}, the charged gluons become tachyonic in the 2SC phase at moderate densities after imposing electrical and color neutralities and
$\beta$ equilibrium conditions, while in the CFL phase the corresponding charged gluons become tachyonic under the previous constraints and at densities where the $s$
quark mass $M_{s}$ becomes a relevant parameter \cite{Fukushima}. Here, I will discuss how the chromomagnetic instabilities in the 2SC system in the absence of an applied magnetic field can be removed by the spontaneous generation of a condensate of inhomogeneous gluons that simultaneously induce a rotated magnetic field. It is expected that a similar mechanism can be also found for the unstable CFL phase, although it is a pending task.

In the gapped $2SC$ phase the solution of the neutrality conditions
$\partial \Omega_{0}/\partial \mu_{i}=0$, with $\mu_{i}=\mu_{e},
\mu_{8}, \mu_{3}$, and gap equation $\partial \Omega_{0}/\partial
\Delta=0$, for the effective potential $\Omega_{0}$ in the
mean-field approximation, led to $\mu_{3}=0$, and nonzero values of
$\mu_{e}$, and $\mu_{8}$, satisfying $\mu_{8}\ll \mu_{e}<\mu$ for a
wide range of parameters \cite{Huang}. Here $\mu$ is the quark
chemical potential, $\mu_{e}$ the electric chemical potential, and
the "chemical potentials" $\mu_{8}$ and $\mu_{3}$ are strictly
speaking condensates of the time components of gauge fields,
$\mu_{8}= (\sqrt{3}g/2)\langle G_{0}^{(8)}\rangle$ and $\mu_{3}=
(g/2)\langle G_{0}^{(3)}\rangle$. The nonzero values of the chemical
potentials produce a mismatch between the Fermi spheres of the quark
Cooper pairs, $\delta \mu=\mu_{e}/2$.

The gapped 2SC turned out to be unstable once the gauge fields
$\{G_{\mu}^{(1)}, G_{\mu}^{(2)}, G_{\mu}^{(3)},$  $K_{\mu},
K_{\mu}^{\dag}, \widetilde{G}^{8}_{\mu}, \widetilde{A}_{\mu}\}$ were
taken into consideration. As shown in Ref. \cite{Igor}, the gluons
$G_{\mu}^{(1,2,3)}$ are massless, the in-medium $8^{th}$-gluon
has positive Meissner square mass, and the $K$-gluon doublet (\ref{charged-fields-2SC})
has Meissner square mass that becomes imaginary for $\Delta > \delta
\mu > \Delta/\sqrt{2}$, signalizing the onset of an unstable ground
state. The mass of the in-medium (rotated) electromagnetic field $\widetilde{A}_{\mu}$
is zero, which is consistent with the remaining unbroken
$\widetilde{U}(1)_{em}$ group. In what follows, we will find an stable ground state solution in the
gapped 2SC phase near the critical point
$\delta\mu_{c}=\Delta/\sqrt{2}$.

To investigate the condensation phenomenon triggered by the
tachyonic modes of the charged gluons, we can restrict our analysis
to the gauge sector of the mean-field effective action
that depends on the charged gluon fields
and rotated electromagnetic field. For a static solution, one only
needs the leading contribution of the polarization operators in the
infrared limit ($p_{0}=0, |\overrightarrow{p}|\rightarrow 0$). Going
through the critical point $\delta \mu = \delta \mu_{c}$ the order
parameter $m_{M}^{2}$ changes sign and varies continuously,
indicating a second-order phase transition. Hence, near the
transition point both the gluon condensate and the induced magnetic
field should be very small and we can neglect their contribution to
the fermion quasiparticle propagators. Under these conditions, the gauge
sector of the effective action can be written as
\begin{eqnarray}
\label{Eff-Act-3} \Gamma_{eff}^{g}& = & \int d^{4}x
\{-\frac{1}{4}(\widetilde{f}_{\mu
\nu})^{2}-\frac{1}{2}|\widetilde{\Pi}_{\mu}K_{\nu}-\widetilde{\Pi}_{\nu}K_{\mu}|^{2}
\nonumber
 \\
& - & [m_{D}^{2} \delta_{\mu 0} \delta_{\nu 0}+
(m_{M}^{2}-\mu_{8}^{2}) \delta_{\mu i} \delta_{\nu i}+
i\widetilde{q}\widetilde{f}_{\mu \nu}] K_{\mu}K_{\nu}^{\dag}\qquad
\nonumber
 \\
 & + &
\frac{g^2}{2}[(K_{\mu})^{2}(K^{\dag}_{\nu})^{2}-(K_{\mu}K^{\dag}_{\mu})^{2}]
+\frac{1}{\lambda}K^{\dag}_{\mu}\widetilde{\Pi}_{\mu}\widetilde{\Pi}_{\nu}K_{\nu}\}\qquad
\end{eqnarray}
where we introduced the 't Hooft gauge with
gauge-fixing parameter $\lambda$, and the notations
$\widetilde{\Pi}_{\mu}=\partial_{\mu}
-i\widetilde{q}\widetilde{A}_{\mu}$ and $\widetilde{f}_{\mu
\nu}=\partial_{\mu}\widetilde{A}_{\nu}-\partial_{\nu}\widetilde{A}_{\mu}$.
In the unstable region close to the transition point the Debye
square mass is
$m^{2}_{D}=\frac{2\alpha_{s}\overline{\mu}^{2}}{\pi}[1+(\frac{2\delta\mu^{2}}{\Delta^{2}})]$,
and the magnetic mass $m_{M}$ is imaginary. As usual in theories
with zero-component gauge-field condensates \cite{Chemical-Pot}, $\mu_{8}$
gives rise to a tachyonic mass contribution, although it has a very
small value, since it is parametrically suppressed by the quark
chemical potential $\mu_{8}\sim \Delta^{2}/\mu$ \cite{Igor}.

At this point let us consider for a moment that we have an external
rotated magnetic field $\widetilde{H}$. In this case the effective
action (\ref{Eff-Act-3}) becomes that of a spin-1 charged field in a
magnetic field (\ref{Eff-Act-2}). If we assume
$\delta \mu < \delta \mu_{c}$, the ground state is stable
considering $m^{2}_{M}-\mu^{2}_{8}> 0$. As discussed in the previous Section, when
$q\widetilde{H}\geq \widetilde{q}\widetilde{H}_{c}=
(m^{2}_{M}-\mu_{8}^{2})$, the magnetic mass of one of the charged
gluon modes becomes imaginary due to the anomalous magnetic moment
term $i\widetilde{q}\widetilde{f}_{\mu \nu}K_{\mu}K_{\nu}^{\dag}$.
This field-induced instability triggers the formation of a
gluon-vortex state characterized by the antiscreening of the
magnetic field.

Now, let us go back to the situation of interest in the present
analysis, that is, a system with no external magnetic field. As
discussed above, if $\delta \mu > \delta \mu_{c}$, the
magnetic mass of the $K$ gluons becomes imaginary:
$(m^{2}_{M}-\mu_{8}^{2})< 0$. Borrowing from the experience gained
in the case with external magnetic field, we expect that this
instability should also be removed through the spontaneous
generation of an inhomogeneous gluon condensate $\langle K_{i}
\rangle$ capable to induce a rotated magnetic field thanks to the
anomalous magnetic moment of the spin-1 charged particles. Having
this in mind, we propose the following ansatz
\begin{eqnarray}  \label{condensate}
\langle K_{\mu} \rangle \equiv \frac{1}{\sqrt{2}}\left(
\begin{array}{cc}
\overline{G}_{\mu}\\
0
\end{array}
\right) \ , \quad \overline{G}_{\mu} \equiv
\overline{G}(x,y)(0,1,-i,0),
\end{eqnarray}
where we took advantage of the $SU(2)_{c}$ symmetry to write the
$\langle K_{i} \rangle$-doublet with only one nonzero component.
Since in this ansatz the inhomogeneity of the gluon condensate is
taken in the $(x,y)$-plane, it follows that the corresponding
induced magnetic field will be aligned in the perpendicular
direction, i.e. along the z-axes, $\langle\widetilde{f}_{12}
\rangle=\widetilde{B}$. The part of the free energy density that
depends on the gauge-field condensates,
$\mathcal{F}_{g}=\mathcal{F}-\mathcal{F}_{n0}$, with
$\mathcal{F}_{n0}=-\Gamma_{0}=\Omega_{0}$ denoting the system
free-energy density in the absence of the gauge-field condensate
($\overline{G}=0, \widetilde{B}=0$), is found, after fixing the
gauge parameter to $\lambda=1$ and using the ansatz
(\ref{condensate}) in (\ref{Eff-Act-3}), to be
\begin{eqnarray}
\label{free-energy} \mathcal{F}_{g} =
\frac{\widetilde{B}^{2}}{2}-2\overline{G}^{\ast}\widetilde{\Pi}^{2}
\overline{G}+2g^{2}|\overline{G}|^{4}\qquad\qquad \nonumber
\\
 -2[2\widetilde{q}\widetilde{B}+(\mu_{3}+\mu_{8})^2+m_{M}^{2}]|\overline{G}|^{2}\qquad
\end{eqnarray}
From the neutrality condition for the $3^{rd}$-color charge it is
found that $\mu_{3}=\mu_{8}$. The fact that $\mu_{3}$ gets a finite
value just after the critical point $m^{2}_{M}-\mu^{2}_{8} = 0$ is
an indication of a first-order phase transition, but since $\mu_{8}$
is parametrically suppressed in the gapped phase by the quark
chemical potential $\mu_{8}\sim \Delta^{2}/\mu$ \cite{Igor}, it will
be a weakly first-order phase transition. Henceforth, we will
consider that $\mu_{3}=\mu_{8}$ in (\ref{free-energy}), and work
close to the transition point $\delta \mu \geq \delta \mu_{c}$
which is the point where $m_{M}^{2}$ continuously changes sign to a
negative value. For very small, negative values of $m_{M}^{2}$, the
gluon condensate and the induced magnetic field should be very small
too, thereby facilitating the calculations.

Minimizing
(\ref{free-energy}) with respect to $\overline{G}^{*}$ gives
\begin{equation}
\label{G-Eq} -\widetilde{\Pi}^{2}
\overline{G}-(2\widetilde{q}\widetilde{B}+| m_M^{2}|)\overline{G}+2g^{2}|\overline{G}|^{2}\overline{G}=0
\end{equation}
Eq. (\ref{G-Eq}) is a highly non-linear differential equation that
can be exactly solved only by numerical methods. Nevertheless, we
can take advantage of working near the transition point, where we
can manage to find an approximated solution that will lead to a
qualitative understanding of the new condensate phase. With this
aim, and guided by the experience with the external field case, where the
solution is always such that the kinetic term
$|\widetilde{\Pi}_{\mu}K_{\nu}-\widetilde{\Pi}_{\nu}K_{\mu}|^{2}$ is
approximately zero near the transition point, we will consider that
when $\delta\mu \simeq \delta\mu_{c}$ our solution will satisfy the
same condition. Hence, we will look for a minimum solution
satisfying
\begin{equation}
\label{G-Eq-2} \widetilde{\Pi}^{2}
\overline{G}+\widetilde{q}\widetilde{B}\overline{G} \simeq 0.
\end{equation}
With the help of (\ref{G-Eq-2}) one can show that the minimum
equation for the induced field $\widetilde{B}$ takes the form
\begin{equation}
\label{B-Eq} 2\widetilde{q} |\overline{G}|^{2}-\widetilde{B}\simeq 0
\end{equation}
The relative sign between the two terms in Eq. (\ref{B-Eq}) implies
that for $|\overline{G}|\neq 0$ a magnetic field $\widetilde{B}$ is
induced. The origin of that possibility can be traced back to the
anomalous magnetic moment term in the action of the charged gluons.
This effect has the same physical root as the paramagnetism found in
Ref. \cite{Vortex} and discussed in the previous Section; where contrary to what occurs in conventional
superconductivity, the resultant in-medium field $\widetilde{B}$
becomes stronger than the applied field $\widetilde{H}$ that triggers
the instability.

Using the minimum equations (\ref{G-Eq}) and (\ref{B-Eq}) in
(\ref{free-energy}), we obtain the condensation free-energy density
\begin{equation}
\label{F-min} \mathcal{\overline{F}}_{g} \simeq
-2(g^2-\widetilde{q}^2)|\overline{G}|^4
\end{equation}
The hierarchy between the strong ($g$) and the electromagnetic
($\widetilde{q}$) couplings implies that
$\mathcal{\overline{F}}_{c}< 0$. Therefore, although the induction
of a magnetic field always costs energy (as can be seen from the
positive first term in (\ref{free-energy})), the field interaction
with the gluon anomalous magnetic moment, produces a sufficiently
large negative contribution to compensate for the increase.
Consequently, as seen from (\ref{F-min}), the net effect of the
proposed condensates is to decrease the system free-energy density.

It follows from Eqs.(\ref{G-Eq})-(\ref{B-Eq}) that near the phase
transition point the inhomogeneity of the condensate solution should
be a small but nonzero correction to a leading constant term
\begin{equation}
\label{Constraint-3} |\overline{G}|^{2}\simeq
\Lambda_{g/\widetilde{q}} |m_{M}^{2}|/2\widetilde{q}^{2} +\mathcal
{O}(m_{M}^4)f(x,y),
\end{equation}
\begin{equation}
\label{Constraint-2} \widetilde{q}\widetilde{B}\simeq
\Lambda_{g/\widetilde{q}} |m_{M}^{2}|+\mathcal {O}(m_{M}^4)g(x,y).
\end{equation}
with
$\Lambda_{g/\widetilde{q}}\equiv(g^{2}/\widetilde{q}^{2}-1)^{-1}$.

The explicit form of the inhomogeneity can be found from
(\ref{G-Eq-2}), which can be written in polar coordinates  as
\begin{equation}
\label{Vortex-Eq} [\frac{1}{r}\partial_{r}(r\partial_{r})
+\frac{1}{r^2}\partial_{\theta}^2+\frac{1}{\xi^2}(1-i\partial_{\theta})-\frac{r^{2}}{4\xi^4}]G(r,\theta)=0
\end{equation}
In the above equation we approached the rotated magnetic field by
its leading in (\ref{Constraint-2}), used the symmetric gauge
$\widetilde{A}_{i}=-(\widetilde{B}/2)\epsilon_{ij}x_{j}$, and
introduced the characteristic length $\xi^{2}\equiv
1/[\Lambda_{g/\widetilde{q}}|m^2_M|]$. Using just the
leading contribution of $B$ in (\ref{Vortex-Eq}) is a consistent
approximation if we simultaneously drop the $\frac{r^{2}}{4\xi^4}$
term and restrict the solution to the domain $r\ll\xi$. Notice that
this domain is indeed a large region because near the critical point
$\xi\gg1$. The most symmetric solution of Eq.(\ref{Vortex-Eq}) is
the one that preserves the SO(2) symmetry in ($x,y$) plane. Hence,
proposing a solution of the form $G(r,\theta)\sim R(r)e^{i\chi}$,
with $\chi$ a constant phase, the equation for $R(r)$ reduces to
$[r\partial_{r}(r\partial_{r}) +\frac{r^{2}}{\xi^2}]R(r)=0$. It is
solved by the Bessel function of the first kind $J_0(r/\xi)$. Then,
the gluon condensate can be written as
$G(r)=(1/\sqrt{2}\widetilde{q}\xi)J_0(r/\xi)\exp i\chi$, which is
consistent with (\ref{Constraint-3}), as in the domain of validity
of this solution ($r\ll \xi$) the series can be approximated by its
first terms. Accordingly, the modulus of the condensate square
is given by
\begin{equation}
\label{Amplitude}
|\overline{G}|^2\simeq\frac{1}{2\widetilde{q}^2\xi^2}-\frac{r^2}{4\widetilde{q}^2\xi^4}
\end{equation}
The improved solution for $\widetilde{B}$ is found substituting
(\ref{Amplitude}) back into (\ref{B-Eq}). The induced field
$\widetilde{B}$ is homogeneous in the $z$-direction and
inhomogeneous in the $(x,y)$-plane.

One may wonder whether this inhomogeneous gluon condensate forms a
vortex state. To answer this question we can compare our results
with the case with external magnetic field reported in \cite{Vortex} and discussed in the previous Section. For this
we should notice that the mathematical problem we have just solved
is formally similar to that where the instability is induced by a
weak external field.  This would be the situation when the $2SC$
system approaches the transition point from the stable side (real
magnetic mass) and the external magnetic field is slightly larger
than the positive mass square $\widetilde{H}\simeq
\widetilde{H}_{c}= m_{M}^{2} \ll 1$. We know that at large
$m_{M}^{2}$ the condensate solution is a crystalline array of
vortex-cells with cell's size $\sim \xi\ll1$. At smaller $m_{M}^{2}$
the lattice structure should remain, but with a larger separation
between cells, since in this case $\xi\gg1$. However, the use of a
linear approximation to solve the equations in this case only allows
to explore the solution inside an individual cell ($r\ll\xi$). This
is the same limitation we have in the linear approach followed in
this Section. Therefore, we expect that when the nonlinear
equations will be solved, the vortex arrangement will be explicitly
manifested.
\section{General Remarks and Astrophysical Considerations}
As we have shown in this talk, in color superconductors magnetic fields tend to be reinforced and even generated. Thus, if a color superconducting state is realized in the core of neutron stars, it should have some implications in the magnetic properties of such compact objects. Taking into account that at the moderate high densities that can exist in the cores of neutron stars the charged gluons Meissner masses decrease from values of order $m_{g}$ to values which are close to zero, and that any magnetic field in that medium with values $\widetilde{H}>m_M^2$ will produce the spontaneous generation of vortices of charged gluons that enhance the existing magnetic field, it becomes natural to expect that color superconductivity can have something to do with the generation of the large magnetic fields observed in some stellar objects as magnetars.

Moreover, if it is accepted the
standard explanation of the origin of the magnetar's large magnetic
field through a
magnetohydrodynamic dynamo mechanism that amplifies a seed magnetic
field due to a rapidly rotating protoneutron star, then it will imply that the rotation should have a spin period $<3 ms$.
Nevertheless, this mechanism cannot explain all the features of the
supernova remnants surrounding these objects
\cite{magnetar-criticism}.

 Now, the gluon
vortices we found in Ref. \cite{Vortex} and discussed here, could produce a magnetic field of the
order of the Meissner mass scale ($m_{g}$), which means with a magnitude in the range $\sim 10^{16}-10^{17} G$. As
discussed in Refs. \cite{Phases}, the possibility of generating a
magnetic field of such a large magnitude in the core of a compact
star without relying on a magneto-hydrodynamics effect, can be an
interesting alternative to address the main criticism
\cite{magnetar-criticism} to the observational conundrum of the
standard magnetar's paradigm \cite{magnetars}. On the other hand, to
have a mechanism that associates the existence of high magnetic
fields to color superconductivity at moderate densities can serve to single out the
magnetars as the most probable astronomical objects for the
realization of this high-dense state of matter.

\bigskip
This work was supported in part by the Office of Nuclear Physics of the Department of Energy under contract DE-FG02-09ER41599.

\end{document}